# A Longitudinal Analysis of Public Discourse on AI Ethics in Education Using Twitter Data

**Akriti Bagale[1], Nafisa Mehjabin [1], Ali Ünlü [2], Aditya Johri [1]**
**[1]Information Sciences and Technology, George Mason University, Fairfax, VA, USA and [2]School of Education and Human Development, University of Virginia, Charlottesville, VA, USA**

**Corresponding author: Akriti Bagale**
**Email: abagale@gmu.edu**

**Abstract**

The rapid integration of artificial intelligence (AI) and generative AI (GenAI) into education presents significant opportunities to enhance teaching and learning, while raising ethical concerns about the responsible use of these technologies in educational settings. Understanding how the public perceives and debates these issues is increasingly important for educators, institutions, and policymakers seeking to integrate AI responsibly and equitably. Social media platforms, where such debates unfold frequently and at scale, offer a valuable lens for capturing large-scale, real-time public reactions to key developments as they emerge. In this study, we analyse five years (2019–2024) of discourse on Twitter (now X) to trace the evolving public conversation around AI ethics in education, paying particular attention to the release of ChatGPT as a pivotal moment that reshaped the nature and tone of that discourse. Using BERT-based topic modelling and SetFit sentiment analysis to identify dominant themes and track sentiment over time, we find that the discourse has been predominantly positive across the observation period, with negative sentiment concentrated around specific ethical controversies. More recently, anxieties about academic integrity and the broader implications of generative AI have come to dominate the conversation. Rather than reflecting a polarized debate, public discourse appears pragmatic and largely receptive to AI integration, though accompanied by growing calls for ethical oversight and institutional accountability. By providing a longitudinal account of public sentiment surrounding AI ethics in education, this study informs educators, institutions, and policymakers an empirically grounded understanding of public expectations, informing the development of responsible, transparent, and equitable approaches to AI integration across educational contexts.

## 1. Introduction

Artificial intelligence (AI) has rapidly transitioned from an experimental technology to a ubiquitous presence in everyday life, sparking a wide range of ethical debates. Although AI technologies have enabled notable advances in efficiency, personalization, and scalability across domains, these advantages are accompanied by unresolved concerns about their unintended consequences, social risks, and ethical concerns. A prominent domain where use of AI has been of concern is education. AI-driven applications that use prediction, recommendation, and personalization have proliferated alongside online learning systems (Holmes et al. 2022b). Although these applications promise efficiency and adaptability, they also raise issues regarding privacy, bias, equity, learning incentives, the sustainability of long-run collective knowledge, and outcomes for at-risk students (Holmes et al. 2022b; Acemoglu, Kong, and Ozdaglar 2026).

Given the prominence of education as a domain in the public sphere, discussions of ethical implications of AI usage extend into broader public debate and social networks, especially platforms such as Twitter. Therefore, they offer a valuable lens to examine these debates (Fu, Zhuang, and Zhang 2023; Hua et al. 2022; Chen et al. 2023; Manovich 2012). Although social networks are limited in the sense that they only capture the voices of their users, they still have an out-sized effect on public opinion and policy discourse, which makes them a critical venue to examine debates around emerging technologies (Chen et al. 2023; Fu, Zhuang, and Zhang 2023; Hua et al. 2022; Liang, Shen, and Fu 2017).

AI in education is not just a technical aspect anymore, but a contested socio-technical artifact, and people debate what it should do, not just what it can do. As (Linderoth, Hulten, and Stenliden 2024) highlights, societal perspectives on educational AI are often overshadowed by technical narratives, reinforcing the need for discourse-level interventions. As the use of GenAI has accelerated across educational settings (Kasneci et al. 2023), public perceptions are of increasing interest to understand how the technology is viewed and identify factors that must be addressed to facilitate broader adoption. This study aims to fill this research gap.

We selected Twitter for our research as it is a commonly used platform and examined data from 2019 to 2024. We operationalized key discourse dynamics, including sentiment, thematic dominance, and amplification. We used topic modelling and sentiment analysis to ensure reproducible and data-driven insights into public perceptions. Overall, we make the following three research contributions:

1. Systematically identifying and analysing the core themes, sentiment distributions, and patterns in Twitter conversations about AI ethics in education using unsupervised machine learning approaches to reveal the structure of public discourse;

2. Linking topic-specific amplification and suppression to societal implications, we show how platform-shaped discourse readjusts public priorities, influences agenda setting, and conditions democratic inclusion in policy formation; and,

3. Examining how various factors shape AI education discourse by comparing Twitter dynamics with framings in mainstream media, educator professional communities, and policy documents, thereby illustrating how different communicative contexts construct competing narratives around educational AI adoption.

## 2. Related Work

### 2.1 Current Discourse on AI Ethics in Education

The focus on the ethics of AI in education began primarily with concern about the use of large educational data that supports the development and use of algorithms (Holmes et al. 2022b; Slade and Prinsloo 2013). Concerns were raised about informed consent and privacy, data interpretation, data management, and other aspects of data use such as the granularity of analysis, e.g., institutional versus individual level. This discourse further resulted in a focus on broader issues, such as power relations, surveillance, and the purpose of education (Slade and Prinsloo 2013).

Researchers have also pointed out ethical issues raised by the systems they built; issues such as equity and diversity, surveillance and consent, identity, and confidentiality (Anderson and Simpson 2007). Some concerns have less to do with technology, but with how it is used and the ethical practices of students, such as cheating and plagiarism (Gearhart 2012). Overall, these concerns highlight that ethics in education is not only about the design of technologies, but also about the human practices surrounding their use (Holmes et al. 2022a).

While initial debates largely focused on issues of data, surveillance, and academic integrity, the rapid rise of generative AI tools such as ChatGPT has expanded the conversation. More recent discussions now tackle both the promises of AI-driven innovation in teaching and learning and the risks and challenges these systems pose for educational practice and policy (Giannakos et al. 2024) (Garc´ıa-Lopez´ et al. 2025).

### 2.2 AI in Education Today: Opportunities and Ethical Dilemmas

The integration of AI into education has been described as both transformative and disruptive. On the one hand, generative AI systems such as ChatGPT have been praised for their potential to personalize learning experiences, support accessibility, provide immediate formative feedback, and reduce teacher workload in repetitive or administrative tasks (Kasneci et al. 2023) (Deng et al. 2025). These findings position AI as a catalyst for inclusive and student-centered learning environments. However, alongside these promises, a parallel body of research shows persistent risks.

Concerns about academic integrity and plagiarism dominate the discourse, with studies warning that unrestricted reliance on generative AI tools could undermine the development of critical thinking, independent problem-solving skills, learning incentives and sustainability of long-term collective knowledge (Tlili et al. 2023; Li et al. 2024; Acemoglu, Kong, and Ozdaglar 2026). Ethical risks, such as algorithmic bias, data privacy, and the opacity of large language models, remain unaddressed within educational contexts (Bender et al. 2021). The workforce implications of AI adoption raise further concerns, particularly regarding the future role of teachers, the deskilling of certain educational tasks, and broader labor market disruptions, and are beginning to receive attention, but remain underexplored (Li et al. 2024).

Studies in today’s context highlight a tension between the optimism surrounding AI’s pedagogical potential and the ethical, social, and institutional risks that accompany its integration. While scholarly and policy debates offer valuable insights into the ethics of AI in education, most of the work on the ethics of AI in education has been largely done by academic researchers or policy analysts. It is essential to examine how these concerns, expectations, and contestations are discussed by public stakeholders. Examining discourse on social networks offers one mechanism to better understand these concerns from a broader perspective.

### 2.3 Social Media Discourse on AI Ethics in Educational Contexts

Social media platforms have become critical spaces for shaping and amplifying debates on AI ethics in education. Twitter, in particular, functions as both a barometer of public sentiment and a site where educators, students, technologists, and policymakers negotiate meaning in real time (Wei et al. 2025). Early studies examining ChatGPTrelated discourse identified dominant themes, such as academic integrity, the erosion of critical thinking skills, and broader ethical challenges, including privacy, bias, and fairness (Taecharungroj 2023; Tlili et al. 2023). More recent analyses employing topic modeling and network analysis have revealed five convergent areas of concern: academic integrity, learning and skills development, limitations of AI capabilities, policy and social implications, and workforce challenges. These studies also identified smaller, but significant issues related to security, operations, and system-level disruption (Li et al. 2024).

The study by (Li et al. 2024) is like o urs and uses Twitter’s Academic API (Dec 1 2022–Mar 31, 2023), RoBERTa sentiment, BERTopic, and NodeXL Social Network Analysis. They document five convergent clusters (1) academic integrity, (2) impact on learning/skills, (3) limitation of capabilities, (4) policy & social concerns, (5) workforce challenges, plus lower-salience operation/security themes.

Despite the growing body of research on AI ethics in education, much of the existing scholarship remains fragmented, either emphasizing traditional concerns around data and surveillance or focusing narrowly on the immediate aftermath of ChatGPT's release. What is missing is a longitudinal perspective that traces the evolution of ethical debates across both academic and public domains. By analyzing social media discourse from 2019 to 2024, this study addresses this gap. It not only captures the shifting concerns of diverse stakeholders but also situates these discussions within broader educational and policy debates. In doing so, the study contributes a more comprehensive understanding of how the promises, risks, and challenges of AI in education are being negotiated in real time, offering insights that are critical for educators, policymakers, and developers seeking to shape responsible and inclusive AI integration in education.

## 3. Methods

Twitter (we will refer to the platform using its former name, as that has been the norm in academic papers) serves as a valuable research tool for scholars investigating various scientific topics (Chen et al. 2023; Fu, Zhuang, and Zhang 2023; Hua et al. 2022) and our selection of Twitter as a data source reflects that. We also used Twitter because it was a dataset that was available to us officially, following their Terms of Service, and met the Institutional Review Board requirements for public datasets at our institution. The study was designed as a longitudinal investigation to capture data at multiple time points of interest, including the pandemic, when online learning took off, and the introduction of ChatGPT, which quickly became one of the most used applications in education.

### 3.1 Data collection and data description

Data was collected using twarc2 tool and Twitter's Academic Research API. The search query consisted of terms such as AI, ChatGPT, LLM, education, ethics, teaching, and learning (see Table 1). These terms were chosen to ensure comprehensive coverage of the discourse. The query targeted English-language tweets from U.S.-based users between January 2019 and November 2024. The detailed search parameters, date, and language are provided in Table 1. 14201 tweets were extracted with their metadata information, including retweet count, reply count, like count, hashtags, and author id, among others.

Table 1: Twitter API search terms

| Conditions | Search |
|---|---|
| API tool | Twitter Search API for academic research |
| Search Date | 2019-01-01 – 2024-11-10 |
| Search terms | ("artificial intelligence" OR "AI" OR "generative ai" OR "chatgpt" OR "llm" OR "large language model" OR "largelanguage model") AND ("education" OR "educator" OR "educate" OR "educates" OR "learning" OR "learner" OR "learns" OR "teach" OR "teaching" OR "ethics" OR "AIED" OR "ethical" OR "ethic") |
| Language | English |
| Country | USA |

### 3.2 Data Analysis

Figure 1 provides an overview of our data collection, preprocessing, and analysis process. Specifically, we used two data analysis techniques to analyse the data, each of which is described in the following.

**Sentiment analysis**: We analysed sentiment in AI and ethics in education discourse using a supervised, binary classification approach. Tweets were labeled as Positive or Negative using a SetFit-based classifier (Tunstall et al. 2022). SetFit is a sentence-transformer fine-tuning framework designed for data-efficient text classification. It leverages semantically rich sentence embeddings while remaining robust to short and informal content, making it well-suited for social media text like Twitter.

A codebook was developed prior to annotation to establish clear criteria for positive and negative sentiment classification. The initial definitions were refined iteratively using the first 20 annotations as a calibration set, allowing ambiguous cases to be resolved and category boundaries to be clarified before full annotation was performed. In the finalized codebook, tweets were labeled as positive where content was primarily informational, descriptive, factual, or procedural, including product launches and tutorials, and as negative where content expressed criticism, concern, or opposition. Two annotators independently labeled an initial pool of 90 tweets. Sentiment labels were normalized to address inconsistencies and then binarized (Negative = 0, Positive = 1). Inter-rater reliability was computed on all 90 tweets, yielding a percent agreement of 91.1% and Cohen's $\kappa$ =0.797, indicating substantial agreement. The annotated dataset was partitioned into training and validation sets using an 80/20 split. A SetFit classifier was then initialized with the pretrained sentence-transformer backbone sentence-transformers/all-mpnet-base-v2 and fine-tuned using the SetFit Trainer with a batch size of 16 over 10 epochs.

The trained SetFit model was evaluated on the 18-instance validation set, yielding an overall accuracy of 89%. The confusion matrix revealed that 6 of 7 negative instances and 10 of 11 positive instances were correctly classified, with only one misclassification in each class. For the negative class, both precision and recall were 0.86 (F1 = 0.86), while the positive class achieved precision and recall of 0.91 (F1 = 0.91), suggesting the model's slightly stronger performance on the majority class. The macro-averaged F1 score of 0.88 and weighted-average F1 of 0.89 collectively indicate that the classifier generalizes well across both sentiment categories despite the modest training sample size, demonstrating the label-efficiency of the SetFit framework for lowresource sentiment classification tasks.

After training, the classifier was applied to the full tweet corpus to generate predicted sentiment labels and associated class probabilities (positive, negative) for each tweet. The resulting predictions were used to construct a temporal sentiment timeline at monthly granularity, using tweet creation timestamps to assign each tweet to a month.

Following classifier inference, predicted labels were aggregated by calendar month and normalized to within month proportions to allow sentiment intensity to be examined independently of tweet volume fluctuations over time. A rolling mean was applied to smooth short-term noise while preserving the underlying temporal trend. Sentiment peaks were identified within the smoothed series using *scipy.signal.findpeaks*, with parameters tuned to retain only substantively meaningful surges above the local baseline. The resulting peak months for both positive and negative sentiment series were then annotated against real-world AI related events occurring at the time and are discussed in the results section.

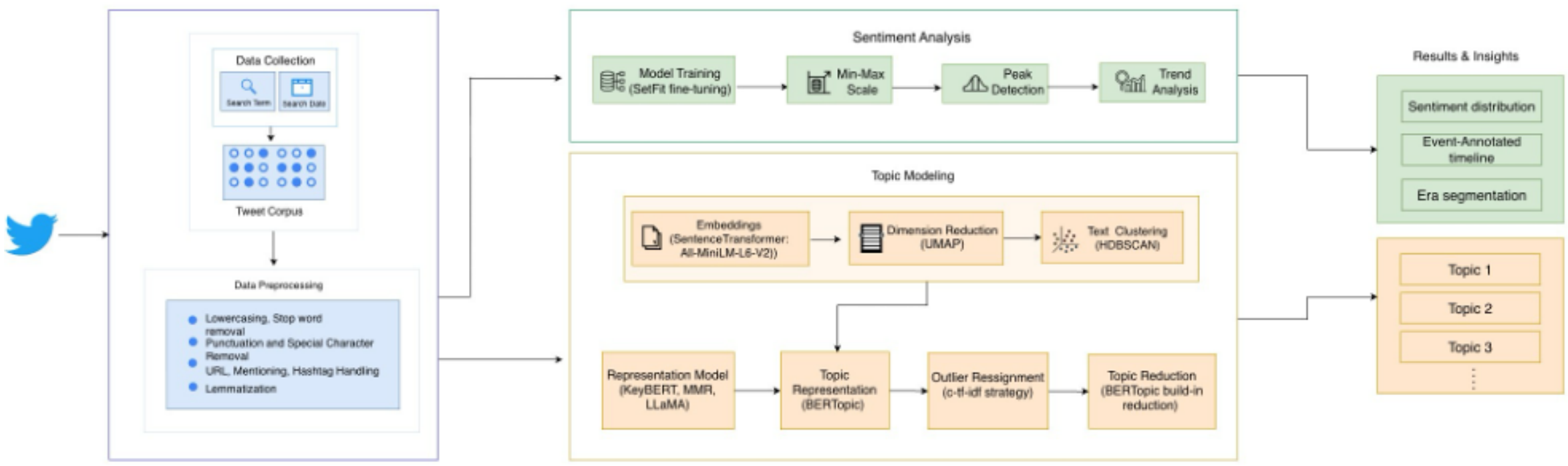

Figure 1: Higher-level workflow of data analysis of AI-Ethics in Education Tweets.

**Bert-based topic modelling:** To gain a thematic understanding of the data, we employed BERT-based topic modeling (BERTopic), a framework that enables interpretable topic extraction using transformer embeddings and clustering (Grootendorst 2022). Because BERTopic combines transformer embeddings, HDBSCAN clustering, and c-TFIDF to produce stable, interpretable topics with auditable diagnostics (e.g., coherence/diversity and transparent term weights), it lets us directly justify that our evaluation is sufficient and aligned with the claims (Grootendorst 2022). Before modeling, the keywords used to extract tweets were treated as stop words to prevent their over-representation in the topics, and a light lemmatization step was applied to normalize word forms while preserving semantic nuance in the discourse.

BERTopic integrates multiple components: sentence embeddings, HDBSCAN, UMAP, and a vectorization model to generate coherent and semantically meaningful topic clusters. For dimensionality reduction, we utilized UMAP (Uniform Manifold Approximation and Projection) (McInnes, Healy, and Melville 2018) with the parameters n neighbors=17, n components=5, min dist=0.0, spread=2.0, and repulsion strength=1.5, random state= 42 and cosine distance as the metric. This setup facilitated a fine-grained separation of document embeddings in a lower dimensional space while preserving local neighborhood structures. The CountVectorizer was applied for feature extraction, using both unigrams and (ngram range = (1, 2)), with min df=5, max df=0.90, and a cap of max features=9000, allowing for a compact and informative vocabulary.

Clustering was performed using HDBSCAN, a hierarchical density-based algorithm (Campello, Moulavi, and Sander 2013), configured with min cluster size=30, min samples=10, and metric='euclidean', to identify dense topic regions while accounting for noise and outliers. For topic modeling, we used the BERTopic framework with the all-MiniLM-L6-v2 Sentence-BERT model for generating embeddings. To enhance topic interpretability, we applied custom representations using KeyBERT, Maximal Marginal Relevance (MMR), and LLaMA-based on generative labeling.

After using BERTopic based on the parameters given, the model yielded 57 topics with 6520(46%) outliers. We addressed the substantial outlier population (6,520 documents, representing 46% of the dataset) using BERTopic's c-TF-IDF strategy. This reassignment successfully reduced outliers to 12 documents (0.08%), ensuring that meaningful content was not excluded from topical analysis. The post-outlier-reassignment model contained 57 topics, exhibiting considerable semantic redundancy as evidenced by hierarchical topic analysis. Inter-topic distance calculations revealed multiple topic clusters with high semantic overlap (cosine similarity 0.8), particularly among AI/machine learning and education-related discourse themes, indicating over-segmentation of the semantic space.

To address this over-segmentation, we applied topic reduction using BERTopic's reducetopics method, specifying nrtopics=16. We selected 16 topics based on the semantic distinctiveness of preserved themes and hierarchical analysis showing successful merging of similar topics (distance scores 0.97-1.64).

To assess the reliability of the manual topic annotation process, two annotators independently labeled a shared set of tweets. Two annotators independently double-coded an initial set of 20 tweets in an iterative piloting phase to refine and clarify the codebook. Inter-rater reliability was then assessed on the remaining double-coded tweets, excluding the pilot set (N = 65). Agreement was high, with a percent agreement of 0.80 and a Cohen's of 0.715, indicating substantial agreement (Landis and Koch 1977) and consistent with IRR estimates reported in comparable annotation tasks on social media data.

**4. Results**

In this section, we discussed our results, starting with sentiment analysis followed by the result of our topic modeling analysis.

**4.1 Sentiment Analysis**

Across the entire corpus of 14,201 tweets, positive sentiment represented 11,595 classifications (81.65%), and negative sentiment comprised the remaining 2,606 (18.35%), as shown in Figure 2. This asymmetry suggests highly optimistic public discourse related to AI during the observation period. Classification confidence was generally high: 95.87% of all predictions fell within the high-confidence subset (0.80), with 11,327 positive and 2,288 negative tweets meeting this threshold.

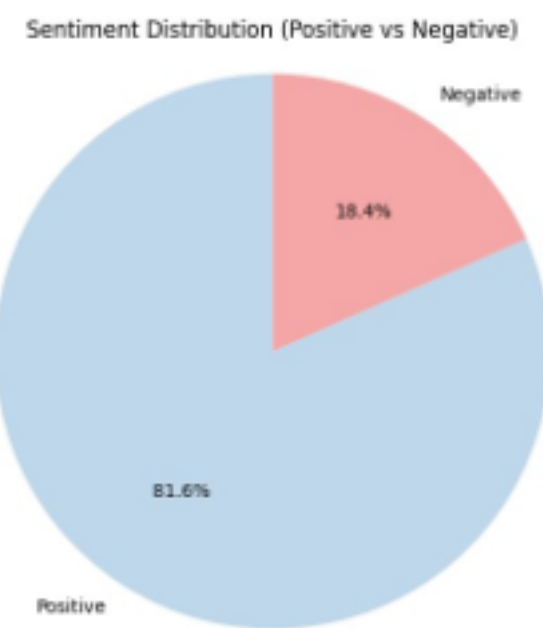

Figure 2: Distribution of Tweet Sentiment Classifications

**4.1.1 Positive Sentiment Peaks**

The positive sentiment curve remained persistently high throughout the study period, which reflects a baseline public enthusiasm for AI in educational and technological contexts. Four positive peaks were identified over the study period, each coinciding with distinct phases of AI development and adoption

**June 2019.** The earliest positive peak occurred with AI discussions at major educational technology conferences. Tweets from this period celebrated AI demonstrations and curriculum integration across K–12 and higher education settings, with practitioners sharing enthusiasm from symposia, showcases, and professional learning events focused on emerging technology in classrooms. The tone was exploratory and aspirational, with educators framing AI not as a threat but as an emerging instructional resource. This peak reflects a moment of early institutional buy-in, when the educational technology community was actively constructing a positive collective narrative around AI adoption ahead of its mainstream arrival in schools and universities.

**May 2021.** A second positive surge was observed with a wave of major cloud AI platform launches, most notably Google Cloud Vertex AI. Tweets celebrated accessible machine learning infrastructure, professional development resources, and expanded educational pathways. The framing positioned AI as both learnable and immediately deployable, lowering the perceived barrier to entry for educators and students alike. This peak aligns with broader trends in cloud adoption and the democratization of ML tooling during the post-pandemic recovery period.

**April 2022.** The third peak reflected the mainstreaming of enterprise AI. Discourse emphasized professional upskilling tweets referenced certifications, competitive AI-powered analytics platforms, and personal learning journeys including university-level AI courses. The sentiment conveyed confident participation, in this phase AI had transitioned from an experimental novelty to a career-defining competency, and the community responded with enthusiasm.

**May 2024.** The final and most pronounced positive peak was shaped by AI's deep integration into everyday educational life. Tweets highlighted Google Gemini for Education, Microsoft and Khan Academy's Khanmigo initiative offering free AI tutoring tools to U.S. teachers, and AI powered personalized learning platforms. High-confidence tweets from this peak consistently framed AI as a complement to human teaching augmenting rather than replacing pedagogical practice, which may account for the particularly strong positive signal observed.

#### 4.1.2 Negative Sentiment Peaks

Unlike the positive peaks, which reflected diffuse adoption enthusiasm, the negative peaks were narrowly clustered around identifiable events, contributing to critical discourse. Four negative peaks were identified, each tightly associated with specific controversies.

**April 2019.** produced the largest sustained negative spike of the early period, almost entirely driven by the collapse of Google's AI Ethics Board. Within days of its formation, the board was dissolved following employee backlash over the inclusion of a conservative think-tank figure and concerns about diversity. Tweets were sharply critical calling out what users saw as performative ethics, institutional hypocrisy, and AI governance failures. Representative tweets included "Google cancels AI ethics board in response to outcry" and "Google dissolves Artificial Intelligence Ethics Board following employee opposition."

**February 2021.** was the second major negative peak, again anchored to Google, specifically the firing of AI ethics researcher Timnit Gebru following her co-authorship of a paper on large language model harms. Tweets expressed outrage at the perceived suppression of internal critique, with several employees publicly resigning in solidarity. The discourse framed this as evidence that corporate AI labs were structurally incapable of honest self-regulation. A recurring theme was the contradiction between Google's "don't be evil" ethos and its treatment of ethics researchers.

**February 2023.** was the largest negative peak after the release of ChatGPT. The release of ChatGPT in late 2022 generated enormous excitement, but by February 2023 the discourse had shifted to its disruptive and ethically fraught implications, particularly for higher education. Tweets raised concerns about academic integrity, students using ChatGPT to pass MBA exams, AI-generated art and copyright, and the model's handling of sensitive content. Noam Chomsky, a prolific philosopher and cognitive scientist, widely circulated the characterization of ChatGPT as "high-tech plagiarism," captured the anxieties of this moment. Unlike the 2019 and 2021 peaks, which were tied to specific institutional events, this negative surge was more diffuse reflecting systemic anxiety about AI's role in knowledge, creativity, and authenticity.

In summary, the sentiment timeline reveals a consistent structural asymmetry: positive peaks were tied to product launches, learning milestones, and expanded access to AI tools, while negative peaks were consistently triggered by failures of AI governance and ethics (2019, 2021) or by anxiety about AI disrupting established educational and creative norms (2023). This asymmetry suggests that the public's emotional relationship with AI in educational and technological contexts is broadly enthusiastic but periodically destabilized by high-profile events that expose gaps between AI's promise and its institutional handling. Importantly, the positive curve did not collapse even as negative events emerged, suggesting a resilient baseline optimism.

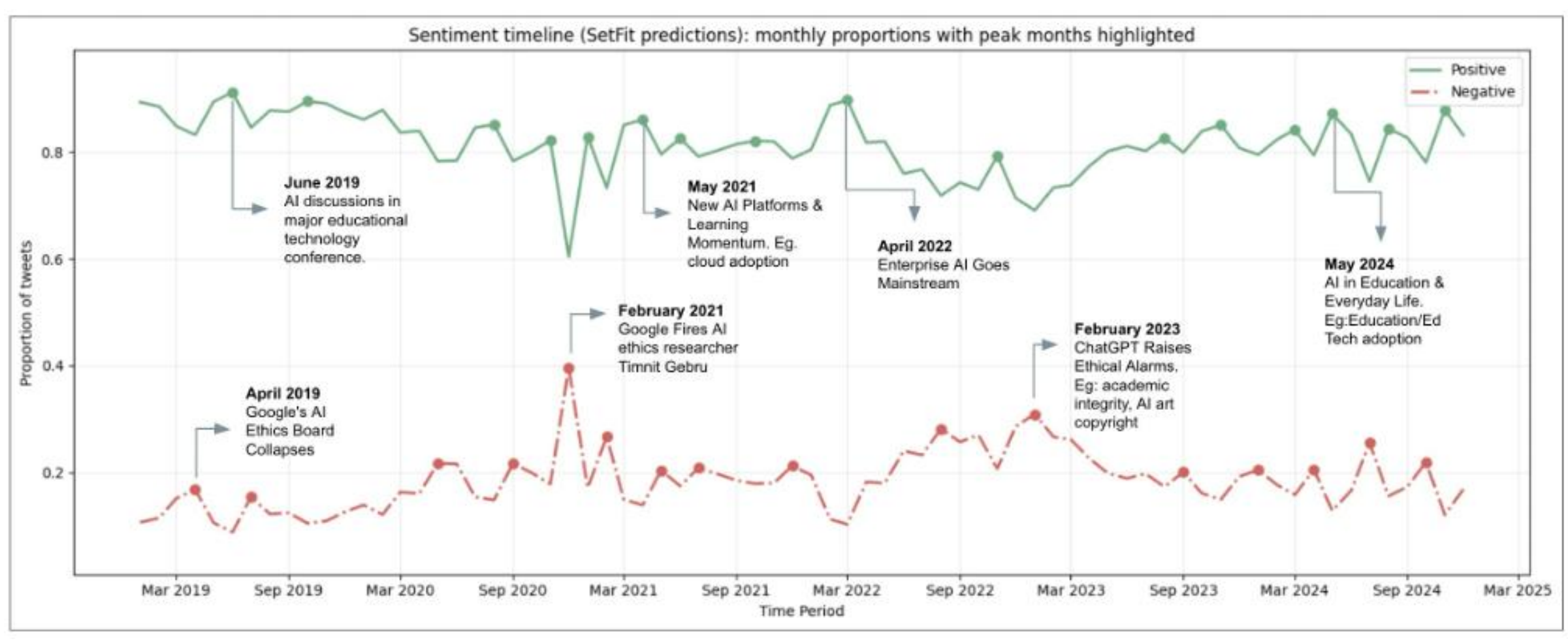

Figure 3: Annotated normalized sentiment timeline. Proportional sentiment fluctuations are mapped with event driven qualitative insights from tweets.

### 4.2 Topic Modeling

The topic modeling, after redistributing the outliers and reducing the topics, yielded 16 topics, but most of them were concentrated in the top 4 topics. Topic 1 was 39.43%, Topic 2 was 21.33%, Topic 3 was 20.83%, and Topic 4 was 7.12%. All other topics were below 3% of the total number of tweets. The remaining outlier after then reduction was 0.084%. The details of the keywords, along with representative tweets and human-assigned topic labels, are shown in Table 3. Due to most of the concentration in the top 4 topics, we decided to investigate the themes within these topics. The themes in the topics are presented below (also see Table 3):

**AI in Education and Pedagogy** Topic 1 was the most prominent in the dataset, comprising 5600 (39.43%) tweets, and represents the core educational discourse surrounding AI integration in learning environments. The predominance indicates substantial public interest in AI's potential in education. Keywords such as student, classroom, teacher, education, learning, edtech, AI, and future-oriented phrases (e.g., "AI classroom", "use AI") indicate a strong inclination to implementation of these tools in educational settings. Representative tweets (see Table 4) predominantly adopt an optimistic stance, envisioning AI as a tool for instructional enhancement rather than systemic disruption. Qualitatively, this topic captures core pedagogical discourse about adopting AI in learning environments rather than replacing human instructors: users frame AI as augmenting instruction (e.g., feedback generation, differentiation) and enabling personalized pathways. The prominence and framing of Topic 1 suggest that the public conversation around AI ethics in education is seen in pragmatic, near-term classroom use cases with a human-in-the-loop ethos.

**Applied Machine Learning and Data Science** Topic 2, consisting of 3,023 tweets (21.33%), captures a broad spectrum of public discourse, regarding educational AI discourse, which suggests cross-domain knowledge transfer and interdisciplinary learning approaches. Dominant Keywords present in this topic were machine, learning, AI, machinelearning, data, datascience, bigdata, analytics, IoT, and Python, with frequent domain markers in healthcare (medical, radiology, healthtech). Qualitatively, this topic captures a broad, technically oriented discourse: tooling, methods, and applied case studies (often health) that cooccur in the education ethics stream. Content reads as continuing education and professional upskilling (e.g., tutorials, examples, workflows) rather than classroom practice. The prevalence of programming languages (Python, JavaScript) and technical frameworks indicates discussions among technically proficient users. This implies that these users are continuing to upskill in the AI/ML domains.

**AI Ethics, Risk & Governance** Topic 3 in the tweet pool includes 2958 (20.83%) tweets and addresses critical ethical considerations in AI deployment, particularly relevant to educational contexts. The terms common in this topic were AI ethics and safety (e.g., ai, ethical, ethic, ai bias, bias, ai safety, aiethics), with frequent references to technology and discussion. Based on the representative tweets and top keywords identified by BERTopic, this topic captures, the discourse centers on identifying and mitigating algorithmic bias, equitable deployment across diverse populations, and responsible data use (privacy and governance). In magnitude, this theme is comparable to the technically oriented Applied ML & Data Science topic 2 (21.33%) and about half the size of the pedagogy-focused AI in Education theme 1 (39.43%), underscoring that ethical considerations are a major but not the only focus of the conversation. Topic 3 suggests that public discussion of AI in education is not confined to classroom practices but substantively engages with risk, fairness, and governance. The representative documents indicate a sophisticated understanding of AI ethics, moving beyond surface-level concerns to nuanced discussions of systemic bias and fairness frameworks.

Table 2: Overall sentiment Distribution of AI-Related Tweets (2019-2024) with Representative Examples

| Sentiment | Count | % | HC (≥0.80) | HC % | Representative Example |
|---|---|---|---|---|---|
| Positive | 11,595 | 81.65 | 11,327 | 97.69 | *"LearnLM is Google's new family of AI models for education"* (May 2024, $p = 0.943$)<br>*"Microsoft teaming up with Khan Academy to give its AI learning tool called Khanmigo free to U.S. teachers"* (May 2024, $p = 0.943$)<br>*"AI is learning how to create itself"* (May 2021, $p = 0.943$) |
| Negative | 2,606 | 18.35 | 2,288 | 87.80 | *"Google cancels AI ethics board in response to outcry"* (Apr 2019, $p = 0.910$)<br>*"Google fires second AI ethics researcher following internal investigation"* (Feb 2021, $p = 0.909$) *"Noam Chomsky on ChatGPT: It's basically hightech plagiarism and a way of avoiding learning"*<br>(Feb 2023, $p = 0.909$) |
| Total | 14,201 | 100.00 | 13,615 | 95.87 | |

*Note.*High-confidence (HC) predictions are defined as those with a predicted probability ≥0.80. HC Percentage refers to the proportion of each sentiment class meeting the high-confidence threshold. Representative examples are selected from peak months based on highest model confidence scores (p). Tweets have been paraphrased for privacy concerns.

**ChatGPT for Teaching, Writing & Coding** Despite its smaller size, topic 3 (7.13%) represents timely and practical concerns about large language model integration in educational settings. The terms appeared in this topic were prompt, ChatGPT/gpt, student, write, code, teach, and learning. Qualitative reading indicates a dual focus: (i) risk/mitigation (cheating, plagiarism detection limits, policy guidance); and (ii) constructive integration (prompt design for drafting, feedback generation, coding support, and lesson planning), with repeated calls for educator training and shareable best practices. Although less prevalent than pedagogy-focused discourse (39.43%), this topic captures a distinct tool-centric layer of the conversation rather than a mere subcomponent of pedagogy: it concentrates on how to use an LLM in prompting strategies, assignment redesign, and management at the classroom level. Its relatively modest size likely reflects the recency of institutional responses and evolving best practices; nevertheless, the content is disproportionately actionable (practice-oriented) compared to the broader ethics and applied-ML themes. The ChatGPT theme operationalizes AI adoption in micropedagogical terms, balancing academic-integrity concerns with classroom-ready practices.

As shown in Figure 4, the graph illustrates the temporal evolution of the top four topics from 2019–2024. Discussions of AI in Education dominate the discourse, peaking sharply in 2023 with the rise of generative AI tools such as ChatGPT, before declining in 2024. Other themes such as Applied ML and Data science, AI Ethics Governance, and ChatGPT for Teaching follow steadier trajectories, with noticeable increases post-2022 as educational applications and ethical debates gained momentum.

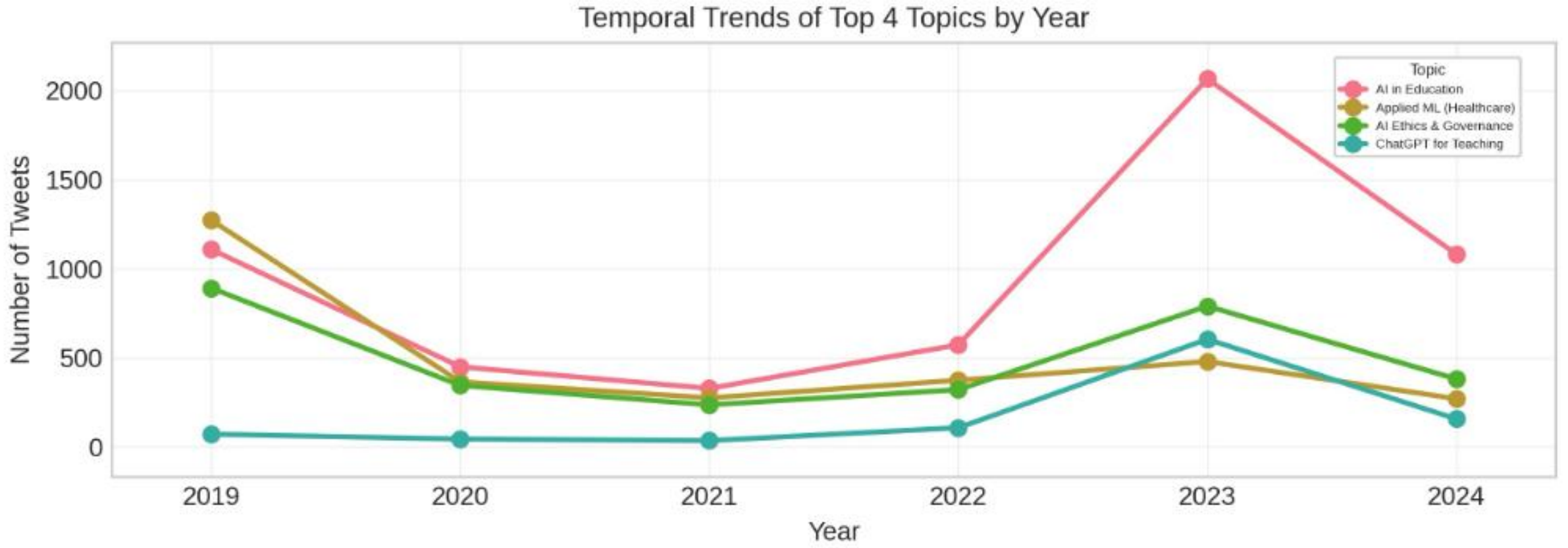

Figure 4: Temporal trends of the top four topics by year.

## 5. Implications and Discussion

In this section, we discuss the implications of the results of our analysis at a broader level to highlight their implications.

### 5.1 Sentiment Reveals the Need for Ethical Integration Rather Than Prohibition

The findings presented here point to a public discourse that has been persistently optimistic about AI, yet marked by moments of concentrated critical concern, a pattern that carries direct implications for how educational institutions should respond to generative AI. The structural dominance of positive sentiment across the observation window, even in the aftermath of high-profile ethics controversies, aligns with technology adoption cycle theory, which suggests that commercial diffusion narratives tend to outpace critical discourse in the public attention economy (Rogers, 2003). This finding aligns with (Tlili et al. 2023), who similarly reported predominantly neutral to positive attitudes toward ChatGPT in education in the early months following its release, suggesting that the initial public reception of generative AI tools has been broadly receptive rather than resistant.

The negative peaks identified in this study, however, complicate any straightforward reading of public optimism. The shift from event-driven negative mobilization, exemplified by the termination of Dr. Timnit Gebru in February 2021, toward the more diffuse anticipatory anxieties observed in February 2023 suggests that public concern had matured beyond reactions to discrete institutional failures and begun to engage with AI as a broader societal issue. Agenda setting theory helps explain this transition: as AI became an increasingly routine feature of public life, the volume and frequency of commercial announcements continuously reset the discursive baseline, meaning that critical concerns had to compete with a growing body of positive framing to gain and sustain public attention (McCombs & Shaw, 1972). The concentrated negative discourse around academic integrity violations, the decline of platforms such as Chegg, and concerns about job displacement reflects this dynamic, consistent with patterns identified by (Li et al. 2024), where public reactions to prominent AI developments reveal enthusiasm and concern operating in parallel rather than in opposition.

For educational institutions, these findings carry a clear practical implication. The persistence of positive sentiment alongside growing structural concern suggests that students and the broader public are not rejecting AI but are looking for frameworks to engage with it responsibly. Contrary to prohibition-based institutional responses, the discourse points toward guided critical engagement, a perspective supported by (Dempere et al. 2023), who argue that integrating AI into teaching practices and updating academic misconduct policies is more effective than enforcing blanket bans. Technology adoption cycle theory further reinforces this position: historically, periods of rapid technology diffusion have required institutions to move beyond reactive governance and develop proactive frameworks that equip users to engage critically with new tools rather than simply restrict access to them (Rogers, 2003). As (Wang and Fan 2025) argue, developing critical digital literacy frameworks that equip students to engage thoughtfully with AI's ethical and societal dimensions represents a more educationally sound response than restriction alone. Given that students are already using generative AI inside and outside the classroom, the central institutional question is not whether to permit its use but how to shape that engagement in ways that foster critical thinking and preparedness for AI's growing role in professional and social life.

### 5.2 Algorithmic Mediation of Educational Discourse and Democratic Participation

Our findings reveal how algorithmic curation on Twitter shapes public discourse around AI ethics in education, with significant implications for democratic participation in educational technology governance. The dominance of Topic 1 (46.7% of discourse) demonstrates how platform algorithms may amplify optimistic, future-oriented narratives about educational AI, potentially creating what Boyd and Crawford term "algorithmic amplification" of certain viewpoints over others (boyd and Crawford 2012).

The substantial engagement with technical discussions (Topic 2, 21.3%) and ethical frameworks (Topic 3, 20.83%) suggests that Twitter's engagement algorithms favour content that demonstrates technical sophistication and moral complexity. This creates a feedback loop where more nuanced, expert-driven content receives greater visibility, potentially excluding voices from educators and students who lack technical AI/ML backgrounds. The under-representation of practical implementation concerns (Topic 4, 8.4%) may reflect algorithmic de-prioritization of mundane institutional challenges in favour of more engaging aspirational or controversial content.

The technical complexity evident across topics creates barriers to meaningful participation in AI education policy discussions. While Topic 3 demonstrates sophisticated ethical reasoning around bias and fairness, the prerequisite technical literacy may limit participation to a narrow demographic of users. This aligns with Winner's (Winner 1980) argument about technological artifacts having political qualities. The very structure of AI discourse on social media may inadvertently exclude key stakeholders (particularly classroom teachers and students) from governance conversations. The prominence of healthcare AI applications (Topic 2) within educational discourse illustrates how social media platforms facilitate knowledge transfer across domains. However, this cross-pollination occurs within algorithmically curated environments that may prioritize sensational or high-stakes applications over nuanced educational contexts (Milli et al. 2025). The integration of medical imaging discussions into educational AI discourse suggests both the democratizing potential of social media knowledge sharing and the risk of inappropriate analogies between high-stakes medical applications and everyday educational tools.

The patterns in the discourse suggest that social media mediated discourse around educational AI can inadvertently reproduce existing inequalities in technological governance. While platforms enable broader participation in AI ethics discussions than traditional policy venues (Bourchas and Gioltzidou 2026), algorithmic curation mechanisms may systematically advantage certain types of contributions over others (Bandy and Diakopoulos 2021). This has direct implications for how educational institutions and policymakers interpret "public opinion" about AI in education, particularly when that opinion is filtered through engagement optimizing algorithms that may not represent the full spectrum of stakeholder concerns.

Table 3: Summary of Topics (BERTopic) for the Tweet Corpus

| Topic | Human-Labeled Name | Top Tokens | Summary | Representative(paraphrased) Tweets | % |
|---|---|---|---|---|---|
| 1 | AI in Education & Pedagogy | human; us; future; teacher; student; teaching | Integrating AI into teaching and learning; pedagogy and classroom practice; educator facing adoption. | A tweet highlighted the challenge that, for educators, there are currently no clear answers about AI in education, framing this uncertainty as an opportunity to broaden awareness of AI's potential in the classroom. | 39.43 |
| 2 | Applied ML & Data Science (esp. healthcare) | machine learning; iot; data science; bigdata; analytics; data | Operational ML and data pipelines with a strong healthcare/medical imaging emphasis. | A news report highlighted medical AI systems, such as Google's lung cancer detector and DeepMind's diagnostic tool, that outperform human specialists in accuracy. | 21.33 |
| 3 | AI Ethics, Risk & Governance | ethic;healthcare; intelligence; need; data; ethical | Discussions of bias, fairness, safety, privacy, and policy/panels around responsible deployment. | A commentator stressed that AI in education must address ethical issues such as privacy, fairness, and equitable access, beyond efficiency gains. | 20.83 |

| 4 | ChatGPT for Teaching, Writing & Coding | use; writing; ChatGPT; teach; student; teaching | Hands-on prompting and classroom/coding tactics for writing support, explanations, and debugging. | A shared article raised the question of ChatGPT's implications for teaching, signalling widespread interest in its educational applications. | 7.13 |
|---|---|---|---|---|---|

*Notes:* Tweets are paraphrased for privacy; semantic content preserved. Percentages show share of the 14,201 tweets covered by Topics 4 (remaining 11.27% across other topics)

### 5.3 Comparative Analysis: Policy vs. Practitioner vs. Media vs. Public Discourse

After the analysis of Twitter data, we compare public opinion with three discursive areas around AI in education ethics: policy documents, practitioner perspectives, news media, and public (Twitter) discourse.

**Policy Documents** National and international policy texts articulate high-level ethical principles for artificial intelligence, but their education specificity is uneven and often underdeveloped. (Schiff 2022) in an analysis of 24 national AI strategies, demonstrates that ethical considerations for AI in education (AIED) are largely absent; instead, most strategies privilege economic competitiveness and innovation agendas over classroom-level concerns. Complementing this, (Adams et al. 2023) synthesizes frameworks advanced by UNESCO, UNICEF, and the World Economic Forum, which consolidate familiar principles such as transparency and fairness while introducing education-adjacent issues, including age-appropriate design and teacher wellbeing. At a broader level, (Jobin, Ienca, and Vayena 2019) map 84 global AI ethics guidelines, identifying convergence around transparency, fairness, and accountability, yet also exposing critical gaps in solidarity and sustainability principles, arguably central to schooling contexts. Within the AIED research community itself, (Holmes et al. 2022b) calls for a domain-specific ethics framework spanning data governance, pedagogy, and equity, underscoring the need for multi-stakeholder responsibility. Taken together, these strands reveal a paradox. Policy and institutional documents are rich in principles, norms, and rights, but they rarely translate into operational guidance for classrooms. While global frameworks emphasize the "what" of ethical AI, they are largely silent on the "how" of everyday educational practice.

**Practitioner Perspectives** Studies examining practitioner and educator perspectives highlight both growing awareness of AI ethics and persistent challenges in operationalizing it. (Pant et al. 2024) in a global survey of 100 AI practitioners, identify barriers across general, technical, and human related domains. Their findings reveal an ambivalence: while practitioners broadly endorse fairness and privacy as normative goals, implementation is constrained by resource limitations, skill gaps, and organizational cultures where ethics work remains compliance-driven rather than integrally embedded in development lifecycles. (Borenstein and Howard 2021) extend this diagnosis by arguing that such limitations reflect structural gaps in education and professional training. They call for embedding ethical reasoning across the full span of AI curricula and practice, emphasizing participatory design that meaningfully incorporates affected stakeholders, as well as interdisciplinary collaboration that positions social scientists, humanists, and domain experts as coproducers of AI systems rather than external auditors. Additionally, (Holmes et al. 2022b) brings these concerns into the AIED research community, synthesizing perspectives from leading scholars to highlight the field's continued ethical deficiency. In response, they propose a layered framework that links high-level values to methodological standards and context-sensitive implementation. Crucially, the framework places ethics not as a discrete stage, but as an iterative, crosscutting dimension spanning design, data curation, modeling, evaluation, and classroom integration while explicitly incorporating the voices of teachers, learners, and other stakeholders. Together, these studies show that practitioner discourse, while attentive to fairness, privacy, and accountability, often struggles with translation into sustainable practice. Ethics emerges as simultaneously valued and fragile: acknowledged as necessary but constrained by structural and institutional realities.

**Media Discourse** Across media ecologies, AI ethics is framed through shifting but recurrent patterns of hype, risk, and governance. (Winkel 2025), in a study of German newspapers, shows that responsibility for AI ethics is framed ideologically, with outlets often simplifying or omitting regulatory complexity in favor of sharper narratives about governance crises or institutional accountability. Nam and Bai (2023), focusing on STEM journals and higher-education magazines, identify a parallel discursive pattern: anxieties about academic integrity, job displacement, and the absence of clear institutional policy dominate coverage, with professional media reinforcing frames already circulating in mainstream outlets. Looking at a longer arc, (Ouchchy, Coin, and Dubljevic 2020) review coverage between 2013 and 2018, documenting a shift from uncritical boosterism to somewhat more balanced reporting. Yet even in this later phase, ethical engagement remained superficial, often reduced to abstract references to fairness or bias without concrete operational discussion. Taken together, these studies suggest that while media discourse has grown more rhetorically balanced, it remains largely episodic, clustered around high-profile events, such as product launches, policy announcements, and scandals, and thin on sustained ethical analysis.

**Public Discourse** Our corpus analysis (2019–2024) reveals that public discourse around AI in education is a positive-leaning tone, oriented toward human-in-the-loop practices, and organized around stable thematic clusters. Sentiment analysis shows a largely neutral distribution, followed by positive and then negative polarity, with shifts tightly coupled to external events. Topic modeling further identifies four dominant themes. First, AI ethics in education and pedagogy center on classroom adoption, instructional integration, and public salience of AIED. Second, applied machine learning and data science reflect a technically focused conversation on the portability of ML/AI knowledge, extending beyond educational settings. Third, ethics, risk, and governance encompass discussions of algorithmic bias, equity across diverse populations, and responsible data use. Here, discourse displays a sophisticated ethical vocabulary, moving beyond surface-level concerns toward nuanced debates over systemic fairness and accountability. Finally, tool-centric practices (e.g., ChatGPT for writing and coding) dominate post-2022 discourse, highlighting both opportunities for classroom integration and anxieties around cheating, plagiarism, and academic integrity. Methodologically, co-occurrence patterns show that tool-centric practices and pedagogical adoption are closely intertwined, while ethics/risk/governance serves as a bridge across all other themes.

Twitter discourse on AI in education diverges from other areas in various ways. Policy documents emphasize broad principles but lack classroom detail (Schiff 2021), while Twitter is practice-oriented, mixing with experimentation with ethical debate (Li et al. 2024). Compared to practitioners, who highlight barriers and compliance-driven ethics (Akgun and Greenhow 2022), Twitter voices a more distributed and often more ambitious vocabulary, though skewed toward technically literate users (Herrera-Pavo et al. 2023). Relative to media, which is episodic and lacks in ethical depth, Twitter is also event-driven but sustains more granular engagement with pedagogy and governance (Chuan, Tsai, and Cho 2019; Schwarz and Unselt 2022).

Across policy, practitioner, media, and public spheres, there is broad agreement on core AI ethics principles, like transparency, fairness, accountability, and shared concern with bias and privacy. Yet they differ in education specificity: policy documents, especially general AI strategies (Schiff 2022; Jobin, Ienca, and Vayena 2019), remain largely education-skeptic, while AIED-focused efforts (Holmes et al. 2022b) provide more contextualization. Practitioners emphasize tractable mitigations (e.g., participatory design) and the public contributes micro-level strategies (e.g., lesson planning, safeguards, prompting). Temporal dynamics also diverge media and public debate spike around launches, scandals, or announcements (Tsimpoukis 2025; Brennen, Howard, and Nielsen 2018), while policy and practitioner texts evolve more slowly (Schiff 2022). The risk framework also varies; accordingly, policy balances rights with innovation (Cihon and Maas 2019), media highlights crises (Schwarz and Unselt 2022), practitioners stress mitigations (Akgun and Greenhow 2022), and the public alternates between integrity concerns and classroom experimentation (Li et al. 2024).

## 6. Limitations and Future Work

While this study offers a multi-label, time-aware analysis of Twitter discourse on AI in education, it is subject to several limitations that qualify its findings. First, the dataset is restricted to Twitter, which may not capture discourse occurring on other platforms (e.g., Reddit, YouTube, or academic forums). The corpus is limited to Twitter/X, whose user base tends to be more technically literate and professionally engaged than the general population, which may not reflect how the broader public perceives AI. The platform's algorithm also prioritizes content that generates high engagement, which may mean that more emotionally charged posts are overrepresented in the data. These factors limit how far the findings can be generalized beyond this platform and audience. Second, our peak detection procedure is sensitive to parameter choices, and alternative settings could yield slightly different temporal patterns. Additionally, contextual annotation of peaks based on hashtags and selected tweets provides qualitative depth but inevitably introduces interpretive subjectivity. Finally, while topic modeling identifies key thematic clusters, emerging technologies, such as ChatGPT, require longer longitudinal analysis to capture discourse maturation and stabilization.

Future research should address these constraints through four main directions. First, cross-platform discourse analysis could validate and extend our findings by comparing Twitter discourse with conversations on other digital and academic venues. Second, analyzing the temporal evolution of ethical concerns as AI technologies mature would illuminate how public debates stabilize, diversify, or fade over time. Third, incorporating geographic and demographic variations in AI education discourse could reveal whose perspectives are amplified or marginalized, and how cultural context shapes ethical framings. Fourth, examining the relationship between online discourse and actual educational policy implementation would clarify whether and how public opinion translates into institutional change. Finally, greater methodological robustness could be achieved by distinguishing between user types (corporate, academic, personal) and incorporating human annotation to reduce misclassification and improve interpretability.

## 7. Conclusion

This study provides a multi-label, time-aware analysis of five years of Twitter discourse on AI ethics in education, revealing a public conversation that is both optimistic about classroom adoption and attentive to systemic risks, such as bias, fairness, and academic integrity. Compared to policy documents that articulate principles without operational guidance, practitioner accounts constrained by compliance logistics, and media coverage that remains episodic and shallow, Twitter discourse contributes distinctive micro pedagogical insights and a more ethically sophisticated vocabulary. At the same time, algorithmic amplification raises concerns about whose voices are most visible, suggesting that teachers, students, and marginalized groups may remain underrepresented. By highlighting how ethical principles are debated, adapted, and enacted in real time, our analysis underscores the value of public discourse as both a complement and a corrective to top-down framings, while pointing to persistent gaps in translating principles into equitable classroom practice.

## Reproducibility

The data and code supporting this study are available from the authors upon request. All analyses were conducted in Google Collab Pro using Python 3 with access to an NVIDIA A100 GPU.

## Declarations

Ethical approval

This study used publicly available Twitter/X data and does not involve direct interaction with human subjects. In accordance with George Mason University's IRB guidelines, research involving publicly accessible social media data does not constitute human subjects research and is therefore exempt from IRB review.

Consent to participate

Not applicable. This study did not involve direct recruitment or interaction with human participants; data were derived from publicly available, de-identified social media posts.

Consent to publish

Not applicable. This manuscript does not include any identifiable individual data. All illustrative examples of social media content included in this manuscript have been paraphrased to protect user privacy.

Funding

This work was supported by the National Science Foundation under Award Numbers 2439460, 2319137, and 1954556, and by the USDA/NIFA under Award Number 202167021-35329.

Data Availability

The datasets analysed during the current study were collected from publicly available posts on X (formerly Twitter). The data and code that support the findings of this study are available from the authors upon reasonable request. All analyses were performed in Google Colab Pro using Python 3, with computational resources provided by an NVIDIA A100 GPU. Detailed information about preprocessing steps, modeling configurations, and analysis workflows can be shared to facilitate reproducibility.

Competing interests

The authors declare that they have no financial or non-financial interests that are directly or indirectly related to the work submitted for publication.

Clinical trial number: Not applicable.

**Acknowledgements:** This work is partly supported by U.S. NSF Award# 2439459, 2439460, 2319137, 1954556, and USDA/NIFA Award# 2021-67021-35329. Any opinions, findings, and conclusions or recommendations expressed in this material are those of the authors and do not necessarily reflect the views of the funding agencies.